\documentclass{article}
\usepackage{amssymb}

\usepackage{amsmath}

\input{tcilatex}

\begin{document}

\title{Entangling Two-Atom Through Cooperative Interaction Under Stimulated
Emission}
\author{Y. Q. Guo\thanks{%
E-mail:yqguothp@yahoo.com.cn}, H. S. Song\thanks{%
E-mail:hssong@dlut.edu.cn}, L. Zhou and X. X. Yi \\
\textit{Department of Physics, Dalian University of Technology,}\\
\textit{Dalian 116024, P. R. China}}
\maketitle

\begin{abstract}
We discuss the generation of two-atom entanglement inside a resonant
microcavity under stimulated emission (STE) interaction. The amount of
entanglement is shown based on different atomic initial state. For each kind
of intial state, we obtain the entanglement period and the entanglement
critical point, which are found to deeply depend on driving field density.
In case of atomic state $\left| ee\right\rangle $, the entanglement can be
induced due to STE. In case of atomic state $\left| eg\right\rangle $, there
is a competition between driving field indued entanglement and STE induced
entanglement. When two atoms are initially in $\left| gg\right\rangle $, we
can find a lumbar region where STE increases the amount and period of
entanglement.

\end{abstract}

\baselineskip=17.3pt \ 


\section{Introduction}

One of the most interesting features of quantum mechanics is the correlation
between pairwise-entangled quantum states of two spatially separated
particles, which is called EPR pairs. Besides the applications of EPR pairs
on investigating the conceptual foundations of quantum mechanics, such as
testing the violation of Bell inequality, a great deal of interesting has
been intensively focused on designing and realizing possible quantum
entangling proposals that can be essential ingredients in quantum
communication \cite{1,2,3} and quantum computation \cite{4}. These EPR pairs
can be formatted in different physical systems such as trapped ions \cite{5}%
, spins in nuclear magnetic resonance \cite{6}, superconductor Josephsen
junctions \cite{7}, Cooper pairs in solid states quantum-dots \cite{8}, and
cavity quantum-electrodynamics systems (CQED) \cite{9}. Among these systems,
CQED system has been deeply studied for entangling two atoms or two modes
field in constructing quantum logic gates \cite{10,11} or quantum memory %
\cite{12}. Alternatively, two atoms can be entangled through contineously
driving by a coherent pump field \cite{11,13}, the assistance of a themal
cavity field \cite{14,15,16}, or even the inducement from the atomic
spontaneous emission \cite{15,17}. Generally, these schemes are effective
since there are some kinds of interaction between atom and atom or atom and
cavity field. While, there is another interaction that is uaually not
included: the stimulated emission (STE, which refers to Einstein B
coefficient \cite{18}) of atoms in a atomic ensemble. The STE of atom
emerges when atom in higher energy level is driven by a polarized photon %
\cite{19}. Especially when atomic population inversion is realized, in a
laser system, STE plays a key role in photon absorb-emission process. In
Ref.[20], taking into account the STE, the authors study the resonance
fluorescence spectrum and present five peaks are formed due to STE. In
solving the resonance fluorescent spectrum, the authors treat the emited
photon as a new driving field acting on both atoms since the emited photon
has the identical character with that of driving photon. In this paper, we
consider a system with this interaction and try to analize the entanglement
character of two-atom. This system is discribed in the first section. In
second section, the measurement of entanglement is presented. And in last
section, some results on two-atom entanglement nature are obtained.

\section{Cooperative Interaction Between atoms}

We consider a system constituted by two two-level atoms located in a
nanocavity and a single mode cavity field. Figure 1 shows the schematic
diagram of this system.

The model is discribed as follow: The cavity is isolated from its
surroundings. Both identical atoms have two internal energy levels: an
excited states $\left| e\right\rangle $ and a ground state $\left|
g\right\rangle $. Either atom can, when it is excited to state $\left|
e\right\rangle $, transit to state $\left| g\right\rangle $ under the
driving of an external photonic field with frequency equals the difference
of energy levels $\left| e\right\rangle $ and $\left| g\right\rangle $ and
emit a polarized photon which is identical with the driving one. Also,
either atom can absorb such polarized photon when it is in state $\left|
g\right\rangle $ and jump to state $\left| e\right\rangle $. The cavity wall
is arranged to be mirrors, so that the emitted photon can be fully reflected
and finally be absorbed by the atoms. That is, besides driving photonic
field $E$, the emited polarized photon from atom 1 can also act as a new
driving photonic field $E^{\prime }$ with repect to atom 2, and \textit{vice
versa}, as if the atoms exchange a photon between them owing to STE, this is
a kind of cooperative interaction. As a result, the whole cavity field is
the sum of two fields $E$ and $E^{\prime }$. The Hamiltonian of the system
in the interaction picture reads \cite{20}

\begin{equation}
H=g_{drv}\sum_{i}(a\sigma _{i}^{+}+a^{+}\sigma _{i}^{-})+g_{stm}\sum_{i<>j} 
\left[ \sigma _{i}^{z}(a\sigma _{j}^{+}+a^{+}\sigma _{j}^{-})+(a\sigma
_{j}^{+}+a^{+}\sigma _{j}^{-})\sigma _{i}^{z}\right]
\end{equation}%
where $a$, $a^{+}$ are eliminate and create operators of driving field, $%
\sigma _{i}^{+}=\left| e\right\rangle _{i}\left\langle g\right| $, $\sigma
_{i}^{-}=\left| g\right\rangle _{i}\left\langle e\right| $ are transition
operators of atom $i$ and $\sigma _{i}^{z}=\frac{1}{2}(\left| e\right\rangle
_{i}\left\langle e\right| -\left| g\right\rangle _{i}\left\langle g\right| )$
is the inversion operator of atom $i$, $g_{drv}$ and $g_{stm}$ represent the
coupling strength between atomic transition $\left| e\right\rangle
\longleftrightarrow \left| g\right\rangle $ and field $E$\ or $E^{\prime }$\
respectively. Generally, $g_{stm}$ is determined by STE coefficient, atomic
density and $g_{drv}$. For convenience, we simply call $g_{stm}$ as STE
coupling strength. The second sum for subscripts $i$ and $j$ proceeds for $%
i,j=1,2$ and $i\neq j$. In the following analysis, we will investigate the
entanglement nature of two-atom sub-system evoluted state under this
interaction.

The evolution of the density matrix of global system with initial state $%
\rho (0)$ is controlled by an unitary operator $\hat{U}(t)=e^{-iHt/\hbar }$,
formally, it is $\rho (t)=\hat{U}(t)\rho (0)\hat{U}^{\dagger }(t)$. Under
the above assumption, two two-level atoms form a $2\otimes 2$ dimensional
Hilbert subspace as

\begin{equation}
\left( H\right) _{a}=\left( H_{1}\right) _{a}\otimes \left( H_{2}\right)
_{a}=\left( 
\begin{array}{cc}
\left| e\right\rangle _{1}\left\langle e\right| & \left| e\right\rangle
_{1}\left\langle g\right| \\ 
\left| g\right\rangle _{1}\left\langle e\right| & \left| g\right\rangle
_{1}\left\langle g\right|%
\end{array}%
\right) \otimes \left( 
\begin{array}{cc}
\left| e\right\rangle _{2}\left\langle e\right| & \left| e\right\rangle
_{2}\left\langle g\right| \\ 
\left| g\right\rangle _{2}\left\langle e\right| & \left| g\right\rangle
_{2}\left\langle g\right|%
\end{array}%
\right) \text{.}
\end{equation}%
By expanding $\hat{U}(t)$ into a combination of Taylor series, we rewrite
the evolution operator matrix in the basis of Equ.2 as following analytical
form

\begin{equation}
\hat{U}(t)=\left( 
\begin{array}{cccc}
2g^{2}a\Theta a^{+}+1 & -iga\Phi & -iga\Phi & 2g^{2}\gamma a\Theta a \\ 
-ig\frac{\sin \Omega t}{\Omega }a^{+} & \frac{1}{2}(\cos \Omega t+1) & \frac{%
1}{2}(\cos \Omega t-1) & -ig\gamma \Phi a \\ 
-ig\frac{\sin \Omega t}{\Omega }a^{+} & \frac{1}{2}(\cos \Omega t-1) & \frac{%
1}{2}(\cos \Omega t+1) & -ig\gamma \Phi a \\ 
2g^{2}\gamma a^{+}\Theta a^{+} & -ig\gamma a^{+}\Phi & -ig\gamma a^{+}\Phi & 
2g^{2}\gamma ^{2}a^{+}\Theta a+1%
\end{array}%
\right)
\end{equation}%
, where, we have set $\Theta =\frac{\cos \Omega t-1}{\Omega }$, $\Phi =\frac{%
\sin \Omega t}{\Omega }$, $\Omega =\{2g^{2}[(\gamma ^{2}+1)a^{+}a+\gamma
^{2}]\}^{\frac{1}{2}}$, $g=$ $g_{drv}+g_{stm}$ and $\gamma =\frac{%
g_{drv}-g_{stm}}{g_{drv}+g_{stm}}$. Note that only when $\gamma =1$ does
this matrix equal that in Ref.[14]. This is obvious because $\gamma =1$
corresponds to a system without STE.

What attracts us is the evolution process of the two-atom sub-system density
matrix which is obtained by tracing over the field variables of system
density matrix $\rho (t)$. We can expect the cooperative interaction can
induce atom-atom entanglement during the evolution.

\section{Measurement of entanglement degree}

Whatever be the initial state of two-atom, the time evolution operator $%
\hat{U}(t)$ would reduce most of the off-diagonal elements of the density
matrix. The resulting two-atom density matrix can be written as

\begin{equation}
\rho (t)=\left( 
\begin{array}{cccc}
A & 0 & 0 & 0 \\ 
0 & B & E & 0 \\ 
0 & E & C & 0 \\ 
0 & 0 & 0 & D%
\end{array}%
\right) \text{.}
\end{equation}

Using the entanglement degree defined by Wootters concurrence simplified
from the entanglement of formation \cite{21}

\begin{equation}
C(\rho )=\max \{0,2\max \{\lambda _{i}\}-\sum_{i}\lambda _{i}\}
\end{equation}%
where $\lambda _{i}$ are the four non-negative square roots of the
eigenvalues of the non-Hermitian matrix $\rho (t)\tilde{\rho}(t)$ with $%
\tilde{\rho}(t)=(\sigma _{y}\otimes \sigma _{y})\rho ^{\ast }(t)(\sigma
_{y}\otimes \sigma _{y})$. We obtain

\begin{equation}
\lambda _{1}=\lambda _{2}=\sqrt{A\cdot D}\text{, }\lambda _{3}=E+\sqrt{%
B\cdot C}\text{, }\lambda _{4}=\left| E-\sqrt{B\cdot C}\right| \text{.}
\end{equation}%
So, the Concurrence of the density must be $C(\rho )=\{0$, $2(\min \{E,\sqrt{%
B\cdot C}\}-\sqrt{A\cdot D})\}$. Then, $\min \{E,B\cdot C\}>A\cdot D$ is the
sufficient and necessary condition for emerging two-atom entanglement. Under
this circumstance, the entanglement degree of two-atom subsystem is

\begin{equation}
C(\rho )=2(\min \{E,\sqrt{B\cdot C}\}-\sqrt{A\cdot D}).
\end{equation}

Alternatively, the two-atom entanglement can be measured by the criteria
defined as the patial transposition proposed by Peres and Horodecki which is
written as $\varepsilon =-2\sum\limits_{i}\mu _{i}$ with $\mu _{i}$
corresponding to the negative eigenvalues of the partial trasposition $\rho
_{a}^{T}(t)$ of density matrix. It has been discussed in Ref. [16] that only
when $E^{2}>A\cdot D$ may the entanglement of two-atom be created. This
criteria for entanglement is equivalent with that of Wootters Concurrence
when $E^{2}\leqslant B\cdot C$ (in fact, the equality is obvious in the
following results). Both criterias are suitable for measuring the
entanglement of arbitrary two qubits system whether the system being pure
state or mixed one.

Here, we use Concurrence as the entanglement measurement.

\section{Atom-Atom entanglement discussion}

We assume the initial cavity driving field is in single mode Fork state $%
\left| n\right\rangle _{f}$ with photon number $n$ presenting the cavity
field density. So that the reduced two-atom sub-system density matrix is

\begin{equation}
\rho _{a}(t)=Tr_{f}\rho (t)=Tr_{f}\left[ \hat{U}(t)\rho (0)\hat{U}^{\dagger
}(t)\right] =\sum\limits_{n}\ _{f}\left\langle n\right| \hat{U}(t)\left|
0\right\rangle _{f}\rho _{a}(0)_{f}\left\langle 0\right| \hat{U}^{\dagger
}(t)\left| n\right\rangle _{f}
\end{equation}%
where $\rho _{a}(0)$ is the initial state of two-atom subsystem. The matrix
element $_{f}\left\langle n\right| \hat{U}(t)\left| 0\right\rangle _{f}$
presents the influence of atomic transition on the cavity mode. In the
following analysis, we will present the entanglement nature of two atoms
under STE when they are initially in different state. For convencience, we
set $g_{drv}\equiv 1$ in following analysis.

Firstly, we consider two atoms are both initialy in their excited state,
that is $\rho _{a}(0)=\left| e\right\rangle _{11}\left\langle e\right|
\otimes \left| e\right\rangle _{22}\left\langle e\right| $. We get, in
Equ.5, $A=\left| U_{11}\right| ^{2}=[1+2(n+1)\frac{\cos g\xi t-1}{\xi ^{2}}%
]^{2}$, $B=C=E=\left| U_{21}\right| ^{2}=(n+1)\frac{\sin ^{2}g\xi t}{\xi ^{2}%
}$, $D=\left| U_{41}\right| ^{2}=4\gamma ^{2}(n+1)(n+2)\frac{(\cos g\xi
t-1)^{2}}{\xi ^{4}}$, where $\xi =\sqrt{2[(\gamma ^{2}+1)(n+1)+\gamma ^{2}]}$%
. It has been be stressed in Ref. [14] that no two-atom entanglement can be
generated when two atoms are initially in $\left| e\right\rangle _{1}\left|
e\right\rangle _{2}$ no matter what state the cavity state is. While, when
STE is included, the result is apparently different. We can find that the
necessary and sufficient condition of generating positive concurrence in
Equ.8 is $0\leqslant \gamma <\sqrt{\frac{n+1}{n+2}}$. That is, there exists
a critical point $\gamma _{0}=\sqrt{\frac{n+1}{n+2}}$ ($g_{ste,crit}=\left( 
\sqrt{n+2}-\sqrt{n+1}\right) ^{2}g_{drv}$) which turns out to be the minimum
value of STE coefficient for generating two-atom entanglement that are
initially in excited states. One of the ways to decrease the critical point
is increasing the density of field. Extremely, for large $n$, this point
tends to zero, which means, in a high field density cavity, even a slight
STE can induce two-atom entanglement. For a vacuum field, this point is
about $0.17g_{drv}$. To show these properties, we plot two-atom entanglement
as a founction of Time-$t$ and $\gamma $ in Figure 1a and Figure 1b with
different driving field density $n$. Both Figures present that the
Concurrence is almost a monotone decreasing founction of $\gamma $. Along $t$
axis, the Concurrence presents a sine-quare-like behavior with periodical
maximum and minimum-zero. While, this period can be changed by alternating $%
\gamma $. Only when $g_{ste}=g_{drv}$ does the Concurrence act as an exactly
sine-quare founction $\sin ^{2}\sqrt{2(n+1)}t$ , where the period presents
as $\frac{\pi }{\sqrt{2(n+1)}}$ which is, for example, exactly $\frac{\pi }{2%
}$ for $n=0$ and $\frac{\sqrt{2}\pi }{4}$ for $n=1$. Elsewhere, the period
of entanglement along $t$ axis is $\frac{2\pi }{\xi }$ for a given $\gamma $%
. It is fascinating that the driving field density $n$ not only determines
the critical value of $g_{stm}$, but also takes great influence on the
entanglement-disentanglement period. Generally, the larger the driving field
density the smaller the entanglement period. Another unapparent character of
two Figures is the peak of the Concurrence for a same $\gamma $ can be
increased by increasing field density except for the range of $\gamma $ very
close to zero. This can be easily understood since $g_{stm,crit}$ can be
increased by increasing $n$. In constructing practical quantum logical
gates, we may need strong and long sustained entanglement, thus, we should
make a suitable choice of controllable physical parameters such as the STE
coefficient and the density of monochromatic driving field.

Secondly, we assume one of the atoms is initially excited and the other has
falled to its ground state, thus the two-atom sub-system initial state is $%
\rho _{s}(0)=\left| e\right\rangle _{11}\left\langle e\right| \otimes \left|
g\right\rangle _{22}\left\langle g\right| $. The resulting two-atom density
matrix elements can be obtained as $A=\left| U_{12}\right| ^{2}=n\frac{\sin
^{2}g\xi t}{\xi ^{2}}$, $B=\left| U_{22}\right| ^{2}=(\cos g\xi t+1)^{2}/4$, 
$C=\left| U_{23}\right| ^{2}=(\cos g\xi t-1)^{2}/4$, $D=\left| U_{42}\right|
^{2}=\gamma ^{2}(n+1)\frac{\sin ^{2}g\xi t}{\xi ^{2}}$, $E=\left|
U_{22}U_{23}\right| =\sin ^{2}g\xi t/4$, where $\xi =\sqrt{2[(\gamma
^{2}+1)n+\gamma ^{2}]}$. We can also find the necessary and sufficient
condition of generating positive entanglement is $\gamma \neq \sqrt{\frac{n}{%
n+1}}$. The critical point is $\gamma _{0}=\sqrt{\frac{n}{n+1}}$ ( $%
g_{ste,crit}=\left( \sqrt{n+1}-\sqrt{n}\right) ^{2}g_{drv}$) which can be
shifted towards zero by increasing $n$ (we will show the reason why $n>0$
must be satisfied in this situation). The entanglement situation is shown in
Figure 2a, 2b, 2c. The whole area can be seperated into two regions: the
region where $0\leqslant \gamma <\gamma _{0}$ and the region where $\gamma
_{0}<\gamma \leqslant 1$. In the first region, the entanglement increases
rapidly with respect to $\gamma $. Especially, in the area that $\gamma
\rightarrow 0$, where two coupling strengthes $g_{drv}$ and $g_{stm}$ are
comparable, the entanglement monotonously reaches its peak. While, it can be
easily proved that this peak can never exceed $0.5$ which is the most
probable maximum value of entanglement. For a given $\gamma $, the time
evolution of entanglement presents periodical loss and revival with a period 
$\frac{\pi }{\xi }$. Obviously, to obtain long time sustained entanglement,
STE should be outstanding and driving field density should not be large.
When $n\rightarrow 0$ and $\gamma \rightarrow 0$, the period tends to 
\textit{infinite}! Under this circumstance, large entanglement can never be
obtained in finite time. In the second region, where STE is very weak, the
coupling between atom and driving field is dominating. The result is similar
with that obtained in Ref.[14]. But the resulting entanglement is much weak
(see Figure 2a) and only emerges when driving field density is small (see
Figure 2b). In both regions, the loss and revival of the entanglement also
shows to be periodical. When $n$ tends to zero, the period, which is
approximately $\frac{\pi }{\sqrt{2}\gamma }$, is only dependent on
difference between two coupling strengthes $\gamma $. It should be stressed
that when driving field density is large, and the first-term interaction in
Equ. 2 can hardly induce entanglement, STE can still generate entanglement.
To sum up, we see there is a competition, which depends on $n$ and $\gamma $%
, between first-term-interaction and second-term-interaction in Equ. 2. The
competition behaves with: which is the domination of entanglement, how about
the shift of the critical point and what is the contrast of entanglement
periods in two regions. While, whatever be the competition, apparent STE can
enhance two-atom entanglement.

\section{Conclusion}

We have discussed the generation of two-atom entanglement inside a resonant
microcavity under an auxiliary interaction-STE. The entanglement when two
atoms are initially in $\left| e\right\rangle _{1}\left| e\right\rangle _{2}$
and $\left| e\right\rangle _{1}\left| g\right\rangle _{2}$ is studied. Some
meaningful results are obtained with the assistance of different STE. The
obtained entanglement is simply determined by a analytic sine-like function
of time and difference between two coefficients. In both cases, we obtained
the critical points of generating entanglement as well as the entanglement
peroid. These two quantities can both be controlled by adjusting the density
of driving field or stimulated emission coefficient in experiment scenario.
We have created two-atom entanglement in first case where there is shown
impossible generating entanglement without STE. In second case, we found a
competition between the interaction with and without STE, while, the
amplitude of entanglement with STE is much larger than that without STE, so
is the period. In third case, a lumbar region is found through raising STE
coefficient. This region is some entanglement place corresponding to a
certain time interval where entanglment can hardly be genenrated without STE.

While, in dealing with this system, we do not take into account the atomic
spontaneous emission which leads to a line width of the emitted photon.
Also, we do not inlude the dissipation of the cavity that has been
considered in many papers \cite{13,22}. Even after including these effects,
the STE could still plays an important role in a collectively excited atomic
ensemble. The entanglement induced by stimulated emission should also be
considered when using atoms to construct quantum logic gate or storage of
photon information. Though we only studied two atoms in a large number of an
atomic ensemble, the results can be extended to a multi-atom system.

{\Huge Figure Captions}

Figure 1: Schematic diagram for two two-level atoms system cooperative
initeraction through STE. Under the drive of a photonic field $E$, one
excited atom can fall to its lower state. The emitted photon acts as a new
field $E^{\prime }$. Another atom may be excited by $E^{\prime }$ and jump
to its higher state. And vice versar. Then they complete a process of
cooperative interaction.

Figure 2a: Two-atom (initially in $\left| e,e\right\rangle $) entanglement
as a founction of $t$ and $\gamma $ for $n=1$.

Figure 2b: Two-atom (initially in $\left| e,e\right\rangle $) entanglement
as a founction of $t$ and $\gamma $ for $n=3$.

Figure 3a: Two-atom (initially in $\left| e,g\right\rangle $) entanglement
as a founction of $t$ and $\gamma $ for $n=3$.

Figure 3b: Two-atom (initially in $\left| e,g\right\rangle $) entanglement
as a founction of $t$ and $\gamma $ for $n=1$.

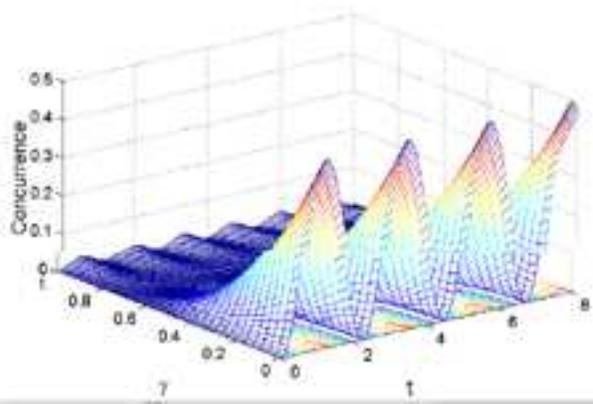
\end{document}